\documentclass[11pt]{article}
\usepackage{fullpage}
\usepackage{amssymb, amsthm, amsmath}
\usepackage{doublespace}
\usepackage{bm}
\usepackage{graphicx}
\usepackage[authoryear]{natbib}
\usepackage{bm}
\usepackage{verbatim}
\usepackage{lineno}
\usepackage{times}

\usepackage[left=1in,top=1in,right=1in]{geometry}
\newcommand{\bbeta}{ \mbox{\boldmath $\beta$}}

\newcommand{\bX}{ \mbox{\bf X}}

\newcommand{\bt}{ \mbox{\bf t}}
\newcommand{\bs}{ \mbox{\bf s}}

\newcommand{\bu}{ \mbox{\bf u}}

\newcommand{\iid}{\stackrel{iid}{\sim}}

\newcommand{\calD}{{\cal D}}
\newcommand{\calS}{{\cal S}}
\newcommand{\calB}{{\cal B}}

\newcommand{\beq}{ \begin{equation}}
\newcommand{\eeq}{ \end{equation}}
\newcommand{\beqn}{ \begin{eqnarray}}
\newcommand{\eeqn}{ \end{eqnarray}}

\begin{document}
\pagestyle{empty}
\begin{center}
{\Large {\bf A spatial capture-recapture model for territorial species}}\\
\vspace{18pt}
{\large \textbf{Brian J. Reich} and
\textbf{Beth Gardner}\footnote{North Carolina State University}}

\vspace{18pt}

\today

\end{center}
\bigskip
\begin{singlespace}
\begin{abstract}

Advances in field techniques have lead to an increase in spatially-referenced capture-recapture data to estimate a species' population size as well as other demographic parameters and patterns of space usage.   Statistical models for these data have assumed that the number of individuals in the population and their spatial locations follow a homogeneous Poisson point process model, which implies that the individuals are uniformly and independently distributed over the spatial domain of interest.  In many applications there is reason to question independence, for example when species display territorial behavior. In this paper, we propose a new statistical model which allows for dependence between locations to account for avoidance or territorial behavior.  We show via a simulation study that accounting for this can improve population size estimates.  The method is illustrated using a case study of small mammal trapping data to estimate avoidance and population density of adult female field voles {\it (Microtus agrestis)} in northern England.

\vspace{12pt}
\end{abstract}
\end{singlespace}

\newpage
\pagestyle{plain}
\setcounter{page}{1}
\begin{center}
{\Large {\bf A spatial capture-recapture model for territorial species}}
\end{center}

\section{Introduction}\label{s:intro}

One essential demographic parameter in making conservation and management decisions is density. To provide reliable estimates of species abundance and density, capture-recapture methods and techniques have been in development for decades \citep{otis_etal:1978, williams_etal:2002}; however, more recently, spatial capture recapture (SCR) models have been developed to explicitly address individual heterogeneity in the capture probability due variable exposure to the trapping \citep{efford:2004, royle_young:2008, borchers_efford:2008, gardner_etal:2009}. In addition to the standard individual encounter histories collected in capture-recapture studies, SCR models make use of the spatial information recorded when traps (or surveys) are spatially replicated.  Thus, information on the individuals' encounter history is supplemented with the locations where the individual has been captured, allowing for estimation of movement of individuals based on these records. A number standard techniques \citep[e.g., small mammal trapping, ][]{converse_etal:2006,ergon_etal:2011,  krebs_etal:2011} use traps placed in array as the typical protocol.  And now, with the development of technologies that allow identification of individuals such as camera traps \citep{oconnell_etal:2010}, hair snares \citep{gardner_etal:2010jwm, sollmann_etal:2012}, acoustic devices \citep{dawson_efford:2009}, scat surveys \citep{thompson_etal:2012} and scent sticks \citep{kery_etal:2010}, more studies are using arrays of devices to monitor populations.  With the increased use of trap arrays, SCR models are becoming more commonly used to estimate demographic parameters and patterns of space usage for a suite of species.  Many advances have been made in SCR modeling in just the past few years including the combination of unmarked individuals \citep{chandler_royle:2013, sollmann_etal:2013}, telemetry information \citep{royle_etal:2013, sollmann_etal:2013}, inhomogeneous point processes \citep{efford_etal:2009euring, efford_fewster:2013, royle_etal:2013}, and open populations \citep{gardner_etal:2010ecol}.

SCR models require a stochastic model for the number of individuals in the population and locations of their home range centers.  Despite recent advances in this area, SCR applications have mostly used a homogeneous Poisson point process model for the home range centers.   This model implies that the individuals are uniformly and independently distributed over the spatial domain of interest. In many applications there is reason to question both of these assumptions, as not all spatial locations will be equally appealing, and many species are known to exhibit territorial or avoidance behavior, particularly during certain seasons of the year, due to mating or denning.  To address variation in density as a function of habitat (or space), the intensity of the point process can be modeled by spatial regions or along a linear gradient \citep{efford_etal:2009euring, efford_fewster:2013}.  This type of inhomogeneous point process, however, does not allow for modeling the interactions between the animals' activity centers, which could provide insight on behavioral activities.  The ability to model such behavior would not only improve density estimation, but would also improve our understanding of the space needs of many species and their distribution on the landscape.  For poorly understood species this could also add novel information about their habitat associations and requirements, particularly important for the long-term management of threatened species.

In this paper, we propose a model that accounts for potential interactions between individuals' activity centers.  We model the locations of the home range locations using a Strauss process \citep{strauss:1975,handbook:2010}, which includes a parameter that determines the strength of repulsion between home ranges.  We show via a simulation study that properly accounting for interactions between individuals can provide a substantial improvement in estimating population size.  For our simulated data generated with interaction, the usual independence model has a significant bias for the population size, and generally has larger uncertainty for the population size than the proposed Strauss process model.

While the Strauss model is intuitive and shows great potential, it presents computational challenges.  First, the likelihood includes a high-dimensional integral that has no closed form.   Extending related work for categorical Markov random fields \citep{green:2002,smith:2006}, we develop an approximation to the Strauss likelihood which allows for posterior sampling.  Second, in our Bayesian analysis, the population size is treated as an unknown parameter to be updated using the data.  As the population size varies, so does the dimension of the likelihood, and thus the posterior.  We overcome this dimension-changing problem using an auxiliary variable scheme in the Markov chain Monte Carlo algorithm.  We present a simulation study to verify that this computational approach leads to reliable inference.  We also present an application of the model to a study of field voles {\it (Microtus agrestis)} in northern England to estimate the population density and we compare these results to one with an homogeneous point process SCR model.   The results suggest that adult female field voles are displaying patterns of avoidance, which results in the posterior mean of density being higher under the Strauss model than the independence model.

\section{A spatial capture-recapture model with interaction}\label{s:model}

The primary objective is to estimate $n$, the number of individuals in the spatial domain of interest $\calD$.  To estimate the population size, counts are recorded for $K$ independent occasions at $J$ traps with spatial locations $\bt_1,...,\bt_J$.  The spatial domain is assumed to be large enough so that it is reasonable to assume that any individual observed in a trap has home range center inside $\calD$.

Let $Y_{ik}\in\{1,...,J+1\}$ be the index of the trap that captured individual $i$ on sampling occasion $k$; if the animal is not captured on occasion $k$, then we set $Y_{ik}=J+1$.   Of course, the challenging aspect of this analysis is that an unknown number of individuals do not appear in the observed data record.   We take the auxiliary variable approach of \cite{royle:2009} to parameterize the unknown population size.  Assume that the prior maximum number of individuals in the population is $N$.  We then parameterize the model using not only the $n$ individuals in the population, but also $N-n$ auxiliary individuals not in the population of interest.  To denote the subset of the individuals that are in the population of interest, let $\delta_i=1$ if individual $i$ is in population and $\delta_i=0$ otherwise, with $\delta_i\iid$ Bern($\pi$).   The population size is estimated using the posterior of $n=\sum_{i=1}^{N}\delta_i$, which has a beta-binomial($N,a_{\pi},b_{\pi}$) prior if $\pi\sim$ Beta$(a_{\pi},b_{\pi})$.

Individual $i=1,...,N$ is assumed to have domain centered on spatial location $\bs_i = (s_{i1},s_{i2})$ and the detection function determined by the function $w_{\rho}$.  Although other functions are possible \citep{efford_etal:2009euring, russell_etal:2012} we use $w_{\rho}(d) = \exp[-0.5(d/\rho)^2]$.  Following \citet{royle_gardner:2011}, the responses for individual $i$ are modeled as
\begin{equation}\label{like}
  \mbox{Prob}(Y_{ik}=j) = \left\{
                            \begin{array}{ll}
                               \frac{\delta_i\lambda w_{\rho}(||\bs-\bt_j||)}{1+\sum_{l=1}^J\delta_i\lambda w_{\rho}(||\bs-\bt_l||)} & k \in \{1,...,J\} \\
                               \frac{1}{1+\sum_{l=1}^J\delta_i\lambda w_{\rho}(||\bs-\bt_l||)} & k = J+1
                            \end{array}
                          \right.
\end{equation}
where $\rho>0$ is the scale of the detection function (which can be related to home range size under this model) and $\lambda>0$ determines the capture rate. Under this specification, Prob$(Y_{ik}=J+1)=1$ for individuals with $\delta_i=0$.

To define the prior for the sampling locations given $\delta$, denote $\calS_1$ as the locations of the $n$ individuals with $\delta_i=1$ and $\calS_0$ as the locations of the $N-n$ individuals with $\delta_i=0$.  The locations of the $N-n$ auxiliary individuals not in the population of interest are arbitrary, and so for convenience we assume the elements of $\calS_0$ are independent and uniform over $\calD$. To account for negative dependence in the domain centers for territorial species, $\calS_1$ are modeled with density corresponding to the Strauss process \citep{strauss:1975, handbook:2010}.  The density is
\begin{equation}\label{strauss}
  p(\calS|a,b) = c_n(a,b)^{-1}\exp\left[-aN(\bs_1,...,\bs_n)\right],
\end{equation}
where $\calS=\{\bs_1,...,\bs_n\}$ and $N(\bs_1,...,\bs_n) = \sum_{i<l}\mbox{I}(||\bs_i-\bs_l||<b)$ is the number of pairs of individuals with home ranges within distance $b$ of each other.   Two parameters control the dependence between locations: $b\ge 0$ controls the distance at which individuals begin to interact and $a\ge 0$ controls the strength of this interaction.  If $a=0$ or $b=0$, then the site locations are independent and uniformly distributed on $\calD$, and if $a>0$ and $b>0$ then locations are repulsed by each other.

The normalizing constant $c_n(a,b)$ is required for $p(\calS|a,b)$ to be a proper density.  It has the form
\begin{equation}\label{c}
  c_n(a,b) = \int_{\calD}...\int_{\calD} \exp\left\{ - aN(\bu_1,...,\bu_n) \right\}d\bu_1...d\bu_n.
\end{equation}
This integral generally does not have a closed form and is required by our computational algorithm if $a$, $b$, or $n$ is unknown.

\section{Computational algorithm}\label{s:comp}

A major computational challenge is evaluating the normalizing constant in (\ref{c}).  To estimate $c$, we adapt a computational method developed for categorical Markov random fields \citep{green:2002,smith:2006}.  The approximation stems from the fact that
\beqn\label{c1}
  \frac{\partial}{\partial a}\log[c_n(a,b)] = -\mbox{E}\left[N(\bs_1,...,\bs_n)|a,b\right],
\eeqn
where the expectation is with respect to the Strauss distribution with parameters $a$ and $b$ for $\bs_1,...,\bs_n$.  Since the locations are independent and uniform when $a=0$, $c_n(0,b) = |\calD|^n$, the area of the spatial domain to the $n^{th}$ power.  Therefore, integrating (\ref{c1}),
\beq\label{c2}
 \log\{c_n(a,b)\} = n\log(|\calD|)-\int_0^a\mbox{E}\left\{N(\bs_1,...,\bs_n)|a',b\right\}da'.
\eeq
This provides a means for estimating $c_n(a,b)$.  We first estimate $\mbox{E}\left\{N(\bs_1,...,\bs_n)|a',b\right\}$ by sampling many sets of locations $\bs_1,...,\bs_n$ from the Strauss model with parameters $a'$ and $b$, and then approximating \\$\mbox{E}\left\{N(\bs_1,...,\bs_n)|a',b\right\}$ as the sample mean of $N$ over the simulations, and standard error as the sample standard deviation divided by $\sqrt{N}$.  This is repeated for several values of $a'$, and $\log\{c_n(a,b)\}$ is approximated by first fitting a tenth-order polynomial function of $a'$ 
via weighted least squares (with weights inversely related to the squared standard errors) and integrating this polynomial function to approximate (\ref{c2}).

We compute $\mbox{E}\left\{N(\bs_1,...,\bs_n)|a',b\right\}$ for $a'$ on the grid 0.0, 0.1, 0.2, ..., 3.0, $b$ on the grid $\calB = \{1,2,...,10\}$, and $n$ on the grid 100, 101, ..., 200.   For each combination of $(a', b, n)$ we use 1,000 samples of $\bs_1,...,\bs_n$, generated using MCMC methods in the {\tt spatstat} package in {\tt R}. A burn-in of 200 MCMC iterations was used for each sample, and the previous draw was used as the initial value for the subsequent sample.  Inspecting the estimates and their associated standard errors suggests this produces reliable estimates. 
While computing these estimates is initially time-consuming, we stress that they are computed a single time, outside any MCMC analysis of a particular dataset, and used for all simulated and real data analysis.  After this initial simulation, evaluating $c$ in the MCMC algorithm is very efficient.

To draw posterior samples, we use the Metropolis within Gibbs algorithm \citep{Chib:1995} implemented in {\tt R}.   Gibbs updates are used for $\delta_i$,  $\pi$, and $b$, which have conjugate full conditionals assuming $b\sim$ Uniform$(\calB)$, where $\calB$ is a discrete grid of values used to compute the normalizing constant.  Their full conditionals are
\beqn
 P(\delta_i=1|\mbox{rest}) &=& \frac{A_1}
                               {A_0+A_1}\nonumber\\
  \pi|\mbox{rest} &\sim & \mbox{Beta}\left(n+a_{\pi},n'-n+b_{\pi}\right)\nonumber\\
  P(b=j|\mbox{rest})& = & \frac{p(\calS_1|a,j)}{\sum_{k\in\calB}p(\calS_1|a,k)}.\nonumber
\eeqn
where $A_0= p(\calS_1^0|a,b)\prod_{k=1}^KP_{ik}(Y_{ik},0)/|\calD|$, $A_1=p(\calS_1^1|a,b)\prod_{k=1}^KP_{ik}(Y_{ik},1)$, $\calS_1^j$ is the set of locations in the population if $\delta_i$ is set to $j$, $P_{ik}(Y_{ik},j)$ is the probability of $Y_{ik}$ in (\ref{like}) given $\delta_i=j$, and  $\calD$ is the area of $\calD$.

We use Metropolis sampling with Gaussian candidate distributions for $a$, $\log(\rho)$, $\log(\lambda)$, and $\bs_i$.  Candidates outside the bounds for $a$ and $\bs_i$ are simply discarded.  The candidate distributions were tuned to give acceptance rate around 0.4.  We generate 50,000 samples and discard the first 10,000 as burn-in.  Convergence is monitored using trace plots and autocorrelations for several representative parameters.

\section{Simulation study}\label{s:sim}

\begin{figure}[h!]
\caption{Spatial domain $\calD$ (shaded gray) and trap locations $\bt_1,...,\bt_J$ (points; measured in meters from the center) for the field voles data.}\label{f:traps}
\begin{center}\begin{picture}(200,200)
 \includegraphics[height=3in,width=3in]{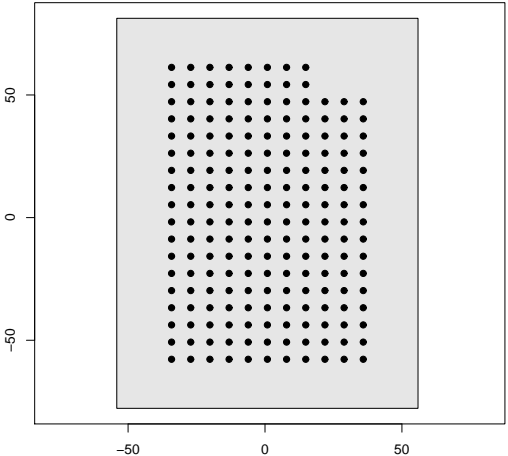}
\end{picture}\end{center}\end{figure}

We conduct a simulation study to validate the performance of our model, and evaluate the performance of the usual homogeneous Poisson process model when its assumptions are violated.   We generate data from the model described in Section \ref{s:model} with spatial domain $\calD$ and $J=192$ trap locations taken from the field voles analysis in Section \ref{s:example} (Figure \ref{f:traps}).  The data are generated to resemble the field voles data, with $n=150$, $N=200$, $\lambda = 0.3$, and $\rho=b=5$.  We consider three simulation designs by varying the Strauss interaction parameter $a$: $a=0$ (no interaction), $a=1$ (moderate interaction), and $a=2$ (strong interaction).  For each of the three values of $a$, we generate $S=100$ data sets.

For each data set, we compare Section \ref{s:model}'s interaction model with the standard no interaction model with $a=0$.  We choose uninformative priors $a_{\pi}=b_{\pi}=1$, $b\sim$ Unif$(\{1,...,10\})$, $a\sim$ Unif$(0,3)$, $\log(\lambda)\sim$ N$(0,1)$, and $\log(\rho)\sim$ N$(2,1)$.   Models are compared using  bias $\frac{1}{S}\sum_{s=1}^S{\hat n}^{(s)}-n$ (``Bias''), mean squared error $\frac{1}{S}\sum_{s=1}^S({\hat n}^{(s)}-n)^2$ (``MSE''), coverage of 90\% intervals $\frac{1}{S}\sum_{s=1}^SI(l^{(s)}\le n \le u^{(s)})$ (``Cover90''), and average width of 90\% intervals $\frac{1}{S}\sum_{s=1}^Su^{(s)}-l^{(s)}$ (``Width90''), where $[l^{(s)},u^{(s)}]$ is the 90\% interval for data set $s$.

\begin{figure}[h!]
\caption{Boxplots of squared errors $({\hat n}^{(s)}-n)^2$ for the 200 simulated datasets for each model and simulation design, and boxplots of the posterior mean of $a$ for the Strauss model for each simulation design (bottom right).}\label{f:sim}
\begin{center}\begin{picture}(270,270)
 \includegraphics[height=4in,width=4in]{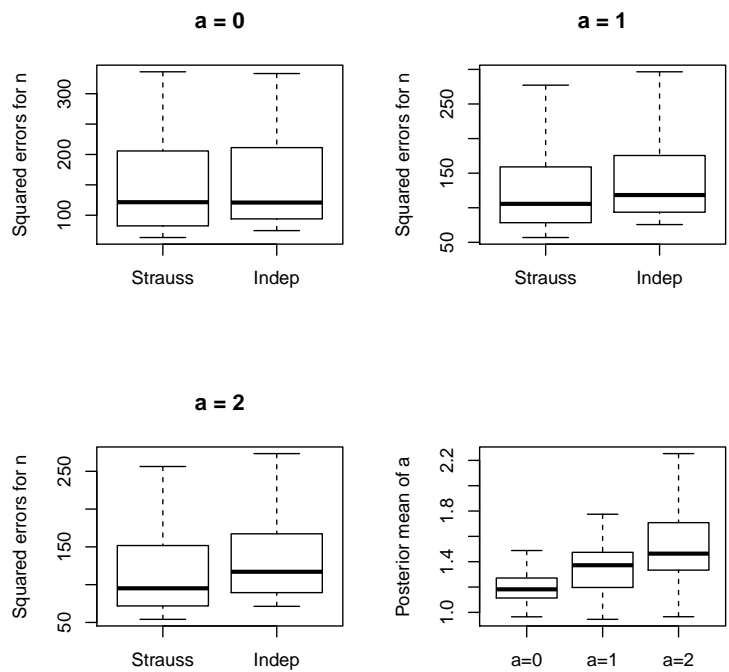}
\end{picture}\end{center}\end{figure}

The results are shown in Table \ref{t:sim} and Figure \ref{f:sim}.  The Strauss model has smaller MSE than the independence model, especially for $a=2$.  The independence model maintains proper coverage but has negative bias for large $a$ and wider posterior intervals than the Strauss model.   Figure \ref{f:sim} shows that the posterior mean of the interaction parameter $a$ is generally higher for datasets with non-zero $a$, but there remains considerable uncertainty about $a$ for these relatively small simulated datasets.

\section{Case Study: Analysis of field voles}\label{s:example}

We fit the model to data collected on field voles ({\it Microtus agrestis}) in Kielder Forest, located on the border between England and Scotland.
The forest comprised of a large spruce plantation (approximately 600 $km^2$), which is mostly grass covered clear-cuts surrounded by dense tree stands.
In this region, field voles are the most numerous of the small rodents.  We selected field voles for this case study as they are known to display territorial or avoidance behavior, with variation by sex, season, density, etc. \citep{pusenius:1993,agrell_etal:1996}.   \cite{ergon_gardner:inreview}  describe the study design in more detail.   A trap array of 196 small mammal traps were placed in a clear cut area, each trap was placed 7 m apart.  Traps were operational for 2.5 days and checked twice day, resulting in 5 secondary sampling occasions. This process was repeated 5 times, approximately once every three weeks, resulting in 5 primary sampling occasions.  We ignored the initial round of sampling, which used only a subset of the traps, and thus we had 4 primary occasions, each with 3 to 5 secondary occasions resulting in a total of 17 occasions. At each occasion, captured animals were marked with ear tags and released; previously captured individuals were recorded.  We used the data from only those individuals that were identified as adult females as their expected behavior of avoidance should be more clearly defined than when males and juveniles are included in the analysis.

The basic model described in Section \ref{s:model} must be modified to allow the population to change by the four primary sampling occasions.  We define $n$ as the number of animals that were present in the population in at least one sampling occasion, and assume that the home range locations $\bs_i$ are constant for all occasions.  For each animal, define $\delta_i$ as an indictor they are included in the population, $\delta_{it}$ as the indicator that they are present in the population at time period $t$, with Prob$(\delta_i=1)=\pi_1$ and Prob$(\delta_{it}=1)=\pi_2\delta_i$. Both $\pi_1$ and $\pi_2$ have Uniform(0,1) priors.  The remaining model is the same as in Section \ref{s:model}.

\begin{figure}[h!]
\caption{Posterior distribution of the population size $n$ under Strauss and independent home range models (top left), conditional mean of $n$ given Strauss parameters $a$ and $b$ for the Strauss model (top right), and the posterior density of $a$ and $(a,b)$ for the Strauss model (bottom row).}\label{f:n}
\begin{center}\begin{picture}(270,270)
 \includegraphics[height=4in,width=4in]{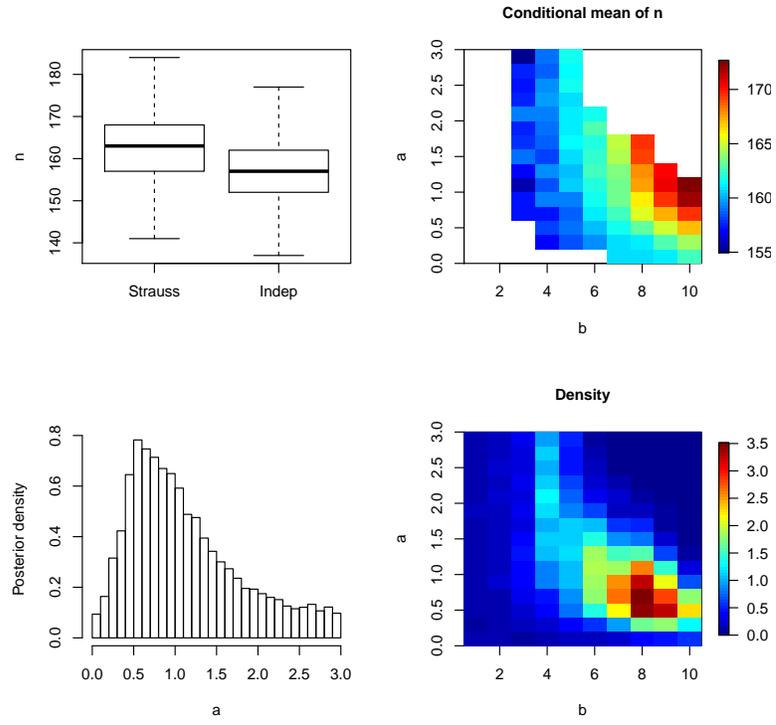}
\end{picture}\end{center}\end{figure}

Table \ref{t:params} and Figure \ref{f:n} compare the population size under the Strauss and independence models.  There is a fairly substantial difference in the posterior of the population size between the fits, with median (90\% interval) 163 (148, 178) for the Strauss model compared to 157 (144, 173) for the independence model.  This agrees with the simulation results, which show that the independence-model estimates are lower on average than the Strauss-model estimate when there is interaction between home range locations.

The posterior of the Strauss parameters in Figure \ref{f:n} shows that the most likely value of the Strauss parameters are $a\approx 0.5$ and $b=8$. There is also posterior mass for a smaller interaction range $b=4$, in which case the strength of interaction increases, with $a>1$ with high probability.   Finally, we note that the posterior 90\% interval of $\pi_2$ is (0.88,0.94), so the population is fairly stable across primary sampling occasion.

\section{Discussion}\label{s:disc}

SCR models are a recently developed and quickly growing class of models.  In the past 4-5 years, the modeling framework has advanced from predominantly a method to estimate density for spatially referenced capture recapture data \citep{borchers_efford:2008, royle_young:2008} to estimating density for unmarked or partially unmarked populations \citep{chandler_royle:2013, sollmann_etal:2013}, evaluating space usage by incorporating telemetry information \citep{royle_etal:2013}, and modeling population dynamics \citep{gardner_etal:2010ecol}.  There has also been work looking at spatial variation in density through the use of an inhomogeneous point process governing the distribution of home range centers \citep{efford_etal:2009euring, royle_etal:2013}.   However, despite the fact that the number of species exhibiting some form or another of territoriality is extensive, ranging from birds, mammals, amphibians, fish, to invertebrates, there has been little to no development of models to account for interactions between home range centers.  To address this gap in SCR models, we proposed a new model for spatially-referenced capture-recapture data that allows for heterogeneity and dependence between home range centers.   To analyze this model, we developed approximate MCMC methods to sample from this complex posterior.  Our simulation study shows that properly accounting for interactions between individuals can substantially improve density estimates.  The simulation study also shows that the model detects the strength and range of interaction of home range centers, thus allowing us to determine patterns of space usages that may arise from behaviors such as avoidance or territoriality in species using spatial capture-recapture data.

In our case study, we estimated the density and space usage patterns of adult female field voles in Northern England.  We found that the home range centers of the voles showed signs of repulsion (hence avoidance) and the range of the process was between 6 and 9 m.  This seems reasonable given that field voles are known to display varying levels of territoriality, with males often defending territories and females changing their behavior during breeding season \citep{pusenius:1993, agrell_etal:1996}.  The size of their home ranges has also been shown to change both with season and density \citep{pusenius:1993}.   Here, we included data over multiple primary sessions because there was not enough data available to model the interactions within a single primary session.  While this limits somewhat our ability to detect patterns of repulsion between individuals since individuals may enter or leave the system between primary occasions, we still found indications of avoidance.

Another issue that may arise in our ability to detect patterns of repulsion is spatial heterogeneity in density.  Often, inhibition between points is only discernible after removing large-scale spatial trends, such as variation in habitat types.   For example, if the spatial domain of interest includes a subregion with unfavorable conditions, then it may appear that individuals are clustering together in the favorable areas, rather than repulsing each other as our territorial model assumes.  Additionally, for some species, like the field vole, habitat quality may lead to variation in home range size and overlap \citep{pusenius:1993}. To model heterogeneity in density, one can use an inhomogeneous point process model for the home range centers \citep{efford_etal:2009euring, royle_etal:2013}.  In our case study, we did not have a spatial covariate available to include in the field vole analysis, but we describe the model in the Appendix for the interested reader.

This work suggests several interesting lines of future research to address computational and biological questions.  One area of future research is to consider alternative computational algorithms to overcome the difficulties posed by the Strauss likelihood.  A natural alternative is the pseudo-likelihood approach, which would approximate the joint distribution of the home range centers as the product of full conditional distributions \citep{handbook:2010}.  This avoids high-dimensional integration, and thus it may be possible to embed this approximate likelihood in the MCMC routine without the initial simulation to approximate the Strauss model's normalizing constant.  Another area of future research is to extend this approach to spatiotemporal data.  SCR models are often used to study changes in population size over time, and this could be accomplished using dynamic modeling of the home range centers allowing also to test the interactions of territoriality, density, and dispersal.

The proposed model allows researchers to examine territoriality (or avoidance behaviors), home range size, and density within one formal modeling framework.  Researchers have studied the interaction of these dynamics for years, looking at the impact of territoriality on resource partitioning and allocation \citep{rubenstein:1981, carpenter:1987, muller_etal:1997},   as well as the influence of patchily distributed habitat on population dynamics of territorial species \citep{winker_etal:1995, johnson_etal:2004}.  These studies often require the live capture or direct observation of species in order to draw inference about these relationships; however, now with the development of  non-invasive techniques, such as camera trapping and scat surveys, we can examine territoriality and its relationship with density and home range size without having to physically capture or observe species.   With the increase in techniques available to collect spatial capture-recapture data, there is much promise for extending SCR models from only estimating density to addressing much broader ecological questions related to space usage, resource allocation, behavior, and movement.

\subsection*{Appendix: Strauss model with spatial covariates}

Let $\bX(\bs)$ be a $p$-vector of spatial covariates, e.g., elevation or land-use classification at location $\bs$, or a polynomial function of $\bs$.  It is possible to include covariates in the Strauss model, $$p(\calS) \propto \exp \left[ \sum_{i=1}^n\bX(\bs_i)^T\bbeta - aN(\bs_1,...,\bs_n)\right],$$ where $\bbeta$ controls the effects of the spatial covariates.  However, the normalizing constant $c$ now depends on $\bbeta$ in addition to $a$ and $b$, which makes the normalizing constant very difficult to compute.

A computationally-convenient alternative is a thinned Strauss process.  As in Section \ref{s:model}, the $n$ locations with $\delta_i=1$ are modeled using the Strauss model and locations with $\delta_i=0$ are modeled as uniform over $\calD$.  To account for covariates, we thin the observations using a second auxiliary variable, $\gamma_i$, whose distribution depends on the spatial covariates, i.e.,
\beqn\label{likeaux2}
  \mbox{Prob}(Y_{ik}=j) &=& \left\{
                            \begin{array}{ll}
                               \frac{\gamma_i\delta_i\lambda w_{\rho}(||\bs-\bt_j||)}{1+\sum_{l=1}^J\gamma_i\delta_i\lambda w_{\rho}(||\bs-\bt_l||)} & k \in \{1,...,J\} \\
                               \frac{1}{1+\sum_{l=1}^J\gamma_i\delta_i\lambda w_{\rho}(||\bs-\bt_l||)} & k = J+1
                            \end{array}
                          \right.\\
  \mbox{P}(\gamma_i=1) &=& \frac{\exp[\bX(\bs_i)^T\bbeta]}{1+\exp[\bX(\bs_i)^T\bbeta]}.\nonumber
\eeqn
Observations are thus included in the population if and only if both $\delta_i=1$ and $\gamma_i=1$, giving $n=\sum_{i=1}^{n'}\gamma_i\delta_i$.

A physical interpretation of this thinned process is that the domain centers with $\delta_i=1$ are the potential domain centers if the entire domain provided conditions favorable to occupancy, and that those with $\gamma_i=0$ are left uninhabited due to unfavorable conditions.  This model is no longer a Strauss process for the $n$ locations in the population with $\delta_i=\gamma_i=1$, however, it gives reasonable limiting cases.  When $a=0$ and there is no interaction between individuals, then this gives the usual thinned representation of the inhomogeneous Poisson process.  On the other hand, when $a=\infty$, then this thinned process and the non-thinned hard-core Strauss process share the property that domain centers within distance $b$ of each other are strictly prohibited.

\section*{Acknowledgments}
We thank Torbj{\o}rn Ergo the field vole dataset and comments on a draft of the manuscript, and all those involved in the data collection.  We also thank Andy Royle for comments on an earlier draft of this manuscript.

\bibliographystyle{rss}
\bibliography{spp_refs}

\newpage

\begin{table}[h!]
\caption{Simulation study results for estimating the population size, $n$.  Monte Carlo standard errors are in parentheses.  The p-value for the paired Wilcoxon test of a difference in MSE is less than 0.001 for all $a$.}\label{t:sim}
\begin{center}\begin{tabular}{lccccc}
Model          & $a$ & MSE & BIAS & Width90 & Cover90  \\\hline
Strauss        & 0 &  157.70 (7.62) &\ 1.23 (0.65) & 27.68 (0.09) & 0.89 (0.02)  \\
Independent    &   &  164.58 (7.41) &\ 0.48 (0.64) & 29.68 (0.10) & 0.93 (0.02)  \\\vspace{-0pt}
&&&&&\\
Strauss        & 1 &  146.79 (8.63) &\ 1.20 (0.63) & 26.95 (0.12) & 0.90 (0.02) \\
Independent    &   &  158.26 (7.66) & -0.94 (0.62) & 29.60 (0.10) & 0.94 (0.02) \\
\vspace{-0pt}
&&&&&\\
Strauss        & 2 &  134.58 (6.80) &\ 1.36 (0.59) & 26.24 (0.10) & 0.87 (0.02)  \\
Independent    &   &  151.07 (7.70) & -1.93 (0.58) & 29.52 (0.10) & 0.95 (0.02)\\
\vspace{-0pt}
&&&&&\\
\end{tabular}\end{center}\end{table}

\newpage

\begin{table}[h!]
\caption{Posterior mean (standard deviation) for model parameters for the Strauss and independence models.}\label{t:params}
\begin{center}\begin{tabular}{lcc}
 & Strauss & Independence  \\\hline
Population size, $n$            &  162.7 (7.69) & 157.4 (7.65)  \\
Scale parameter, $\rho$         &\  5.67 (0.12) &\ 5.67 (0.12)  \\
Baseline detection, $\lambda$   &\  0.25 (0.02) &\ 0.25 (0.02)  \\
Interaction range, $b$          &\  6.45 (2.23) &\ --   \\
Interaction strength, $a$       &\  1.13 (0.68) &\ --   \\
Inclusion probability, $\pi_1$  &\  0.81 (0.05) &\ 0.78 (0.05)  \\
Inclusion probability, $\pi_2$  &\  0.91 (0.02) &\ 0.91 (0.02)
\end{tabular}\end{center}\end{table}

\newpage

\listoffigures

\newpage

\newpage

\newpage

\end{document}